# Frequency transitions in phononic four-wave mixing

Authors: Adarsh Ganesan[1], Cuong Do[1], Ashwin Seshia[1]

[1.] Nanoscience Centre, University of Cambridge, Cambridge, UK

**This work builds upon the recent demonstration of a phononic four-wave mixing (FWM) pathway mediated by parametric resonance. In such a process, drive tones $f_{d1}$ and $f_{d2}$ associated with a specific phonon mode interact such that one of the drive tones also parametrically excites a second mode at a sub-harmonic frequency and such interactions result in a frequency comb $\frac{f_{d1}}{2} \pm n(f_{d1} - f_{d2})$. However, the specific behaviour associated with the case where both drive tones can independently excite the sub-harmonic phonon mode has not been studied or previously described. While it may be possible to expect the merger of two frequency combs $\frac{f_{d1}}{2} \pm n(f_{d1} - f_{d2})$ and $\frac{f_{d2}}{2} \pm n(f_{d1} - f_{d2})$, this paper indicates that only one of these mechanisms is selected and also shows an interesting transition linked to this process. Such frequency transitions from $\frac{f_{d1}}{2} \pm n(f_{d1} - f_{d2})$ to $\frac{f_{d2}}{2} \pm n(f_{d1} - f_{d2})$ holds promise for computing applications.**

Four-Wave Mixing (FWM) has been widely studied in nonlinear physics and has been experimentally observed in wide-ranging physical systems including optics [1-2], atom optics [3-4] and electrical systems [5]. Recently, a specific FWM pathway based on parametric resonance [6] has been identified in a micromechanical resonator. In this mechanism, two tones with frequencies $f_{d1}$ and $f_{d2}$ are involved as inputs. While both of these drive tones directly excite a specific phonon mode, one of the drive tones also parametrically modulates another mode at twice its frequency. This parametric modulation results in an additional tone with frequency of either $\frac{f_{d1}}{2}$ or $\frac{f_{d2}}{2}$. Through higher order interactions of this sub-harmonic tone with $f_{d1}$ and $f_{d2}$, the near resonant tones about either $f_{d1}$ & $\frac{f_{d1}}{2}$ or $f_{d2}$ & $\frac{f_{d2}}{2}$ are generated. In other words, either the tones with frequencies: $f_{d1} \pm n(f_{d1} - f_{d2}); \frac{f_{d1}}{2} \pm n(f_{d1} - f_{d2}); n \in Z$ or $f_{d2} \pm n(f_{d1} - f_{d2}); \frac{f_{d2}}{2} \pm n(f_{d1} - f_{d2}); n \in Z$ are generated. However, the case when both of the drive tones can independently trigger parametric excitation is not discussed. This paper reports the experimental results corresponding to this specific regime. In this manner, evidence of an interesting frequency transition from $\frac{f_{d1}}{2} \pm n(f_{d1} - f_{d2})$ to $\frac{f_{d2}}{2} \pm n(f_{d1} - f_{d2})$ is provided.

To study the frequency transition in the context of other recent demonstrations [6-7], we consider the following dynamics.

$$\ddot{Q}_1 = -\omega_1^2 Q_1 - 2\zeta_1\omega_1\dot{Q}_1 + \alpha_{11}Q_1^2 + \beta_{111}Q_1^3 + \beta_{122}Q_1Q_2^2 + \alpha_{12}Q_1Q_2 + P$$
$$\ddot{Q}_2 = -\omega_2^2 Q_2 - 2\zeta_2\omega_2\dot{Q}_2 + \alpha_{21}Q_1Q_2 + \alpha_{22}Q_2^2 + \beta_{112}Q_1^2 Q_2 \tag{1}$$

where $P$ is the drive, $\alpha_{ij}$ & $\beta_{ijk}$ are quadratic and cubic coupling coefficients and $\omega_{i=1,2}$ and $\zeta_{i=1,2}$ are natural frequencies and damping coefficients of modes $i = 1,2$ respectively.

This dynamics was originally considered to model parametric resonance [8]. Based on this, the excitation of a tone $\frac{f_d}{2}$ can be expected through parametric modulation provided by the term $Q_1Q_2$. However, the recent experiments [7] have shown an interesting possibility to excite tone $\frac{f_0}{2}$ instead of $\frac{f_d}{2}$ when $f_d$ when $f_d$ set outside the dispersion band of a specific phonon mode. Following this excitation, through the cascade of nonlinear interactions, a comb of spectral lines with frequencies $f_0 \pm n(f_d - f_0); \frac{f_0}{2} \pm n(f_d - f_0); n \in Z$ is formed. To further explore the nature of frequency comb, two drive tones: $f_{d1}$ and $f_{d2}$ were presented to this system and this revealed an emergent FWM process [6]. Here, the interactions between the drive tones start to take place when one of the drive tones parametrically modulates another phonon mode at twice its frequency. It is important to note here that the drive tones do not interact in the absence of parametric instability. Similar to three-wave mixing, these interactions also result in a frequency comb but with spacing $|f_{d1} - f_{d2}|$. This specific aspect can also be explained through the same dynamics. Now, we present a more complex case. Here, both of the drive tones can independently undergo parametric instabilities. Based on simple superposition of two independent FWM pathways, we may expect a merger of frequency combs i.e. $f_{d1} \pm n(f_{d1} - f_{d2}); f_{d2} \pm n(f_{d1} - f_{d2}); n \in Z$ and $\frac{f_{d1}}{2} \pm n(f_{d1} - f_{d2}); \frac{f_{d2}}{2} \pm n(f_{d1} - f_{d2}); n \in Z$. While $f_{d1} \pm n(f_{d1} - f_{d2})$ and $f_{d2} \pm n(f_{d1} - f_{d2})$ correspond to the same, $\frac{f_{d1}}{2} \pm n(f_{d1} - f_{d2})$ and $\frac{f_{d2}}{2} \pm n(f_{d1} - f_{d2})$ are separated by $\frac{(f_{d1}-f_{d2})}{2}$. Hence, the spectral lines of frequencies $f_{d1} \pm n(f_{d1} - f_{d2})$ and $\frac{f_{d1}}{2} \pm n\frac{(f_{d1}-f_{d2})}{2}$ are expected. However, our experiments presented in the latter part of this paper indicate otherwise. In contrast to the merger, the excitation of only one of the frequency combs is observed i.e. either $f_{d1} \pm n(f_{d1} - f_{d2}); \frac{f_{d1}}{2} \pm n(f_{d1} - f_{d2}); n \in Z$ or $f_{d2} \pm n(f_{d1} - f_{d2}); \frac{f_{d2}}{2} \pm n(f_{d1} - f_{d2}); n \in Z$. This indirectly suggests an existence of competition between the drive tones $f_{d1}$ and $f_{d2}$ towards the specific formation of frequency combs. Such a nature of nonlinear resonance is also interesting like the three-wave mixing (TWM) and simple four-wave mixing (SFWM) pathways and can be explained using the same dynamics (eq. (1)).

To probe this interesting frequency transition from $\frac{f_{d1}}{2} \pm n(f_{d1} - f_{d2})$ to $\frac{f_{d2}}{2} \pm n(f_{d1} - f_{d2})$ in FWM, the experimental system that was used to study TWM and SFWM is once again utilized. The device consists of an AlN thin film coated Si microstructure of dimensions $1100 \times 350 \times 11 \; \mu m^3$ in the form of a free-free beam topology. On top of the AlN layer, an Al metal electrode is also patterned for applying the electrical signal through the device. On the application of an electrical signal, mechanical deformation is induced in the beam through the piezoelectric effect. The rigidly linked Si and Al layers, along with the AlN layer, vibrate in unison. This vibratory motion is imaged and analysed using Laser Doppler Vibrometry. The experiments were carried out under ambient pressure and temperature conditions.

When an electrical signal of $S(f_{d1} = 3.86 \; MHz)$ is applied, the excitation of phonon mode with frequency closer to $f_{d1}$ is expected. In our experiments, this specific phonon mode corresponds to a length extensional mode. Additionally, for large drive levels, self-excitation of another phonon mode at $\frac{f_{d1}}{2}$ may also be expected via the phenomenon of parametric resonance. The same is the case for $S(f_{d2} = 3.864 \; MHz)$. A similar trend was also captured by our experiments. For $S(f_{d1} = 3.860 \; MHz) = 15 \; dBm$, the excitation of tone at $\frac{f_{d1}}{2} = 1.930 \; MHz$ was observed (Figure 1B). Similarly, for $S(f_{d2} = 3.864 \; MHz) = 15 \; dBm$, the excitation of tone at $\frac{f_{d2}}{2} = 1.932 \; MHz$ was observed (Figure 1C). Now, both of these high intense tones are applied through the separate split electrodes of the device as shown in the figure 1A. Surprisingly, figure 1D showed a frequency comb formation $f_{d2} \pm n(f_{d1} - f_{d2}); \frac{f_{d2}}{2} \pm n(f_{d1} - f_{d2}); n \in Z$. Particularly, the comb about sub-harmonic frequency $\frac{f_{d2}}{2}$ reveals an interesting characteristic. As it can be noted, the tone at $\frac{f_{d1}}{2}$ is missing in such frequency combs and the disappearance of this tone can be linked to a possible FWM pathway as discussed in the theory section.

To systematically understand this anomaly, experiments were carried out at different drive levels of $f_{d1}$ and $f_{d2}$ and the corresponding frequency spectra about $1.9 - 1.96 \; MHz$ were recorded. For the experimental control, $S(f_{d1})$ and $S(f_{d2})$ are individually fed at different drive levels at first. Figures 2A and 2B indicate the parametric thresholds corresponding to $f_{d1}$ and $f_{d2}$ and are $-10 \; dBm$ and $10 \; dBm$ respectively. Now, both $S(f_{d1})$ and $S(f_{d2})$ are fed together. Specifically, $S(f_{d1})$ is kept constant at $10 \; dBm$ and $S(f_{d2})$ is varied from $-15 \; dBm$ to $20 \; dBm$. As seen in figure 2C, the frequency comb of $\frac{f_{d1}}{2} \pm n(f_{d1} - f_{d2})$ is observed for the drive levels $S(f_{d2})$ ranging from $-15$ to $9 \; dBm$. However, when the drive level $S(f_{d2})$ is further increased further, the transition from $\frac{f_{d1}}{2} \pm n(f_{d1} - f_{d2})$ to $\frac{f_{d2}}{2} \pm n(f_{d1} - f_{d2})$ is evidenced. It is important to note here that despite the continued significant excitation of $f_{d1}$, this cross-over takes place. There is certainly a competition between two individual nonlinear pathways. And, specifically, the nonlinear FWM associated with drive tone $f_{d2}$ succeeds when its drive level $S(f_{d2})$ is increased to $13\ dBm$ for $S(f_{d1}) = 10\ dBm$. Additionally, between these two frequency comb regimes, there also exists a transition phase where the frequency comb is not well-defined.

For quantitative assessment, the displacement amplitudes corresponding to the frequencies $\frac{f_{d1}}{2} = 1.93\ MHz; \frac{f_{d1}}{2} + (f_{d1} - f_{d2}) = 1.934\ MHz; \frac{f_{d1}}{2} - (f_{d1} - f_{d2}) = 1.926\ MHz; \frac{f_{d1}}{2} + 2(f_{d1} - f_{d2}) = 1.938\ MHz; \frac{f_{d1}}{2} - 2(f_{d1} - f_{d2}) = 1.922\ MHz; \frac{f_{d2}}{2} = 1.932\ MHz; \frac{f_{d2}}{2} + (f_{d1} - f_{d2}) = 1.936\ MHz; \frac{f_{d2}}{2} - (f_{d1} - f_{d2}) = 1.928\ MHz; \quad \frac{f_{d2}}{2} + 2(f_{d1} - f_{d2}) = 1.940\ MHz$ and $\frac{f_{d2}}{2} - 2(f_{d1} - f_{d2}) = 1.924\ MHz$ are recorded. In R1, the frequency comb corresponding to the frequencies $\frac{f_{d1}}{2} \pm n(f_{d1} - f_{d2})$ is only formed. Hence, the displacement amplitudes at the frequencies $\frac{f_{d2}}{2} \pm n(f_{d1} - f_{d2}); n = 0,1\ \&\ 2$ are minimal. However, after the transition to R3 through R2, the displacement amplitudes at $\frac{f_{d2}}{2} \pm n(f_{d1} - f_{d2}); n = 0,1\ \&\ 2$ build up and the displacement amplitudes at $\frac{f_{d1}}{2} \pm n(f_{d1} - f_{d2}); n = 0,1\ \&\ 2$ diminishes to a value corresponding to the noise floor. Within R1, a continuous increase in the displacement amplitudes at the sidebands $\frac{f_{d1}}{2} \pm n(f_{d1} - f_{d2}); n = 1,2$ is observed corresponding to the increasing drive levels of $S(f_{d2})$. This is due to the higher intensity of drive tone $f_{d2}$. However, in R3, the displacement amplitudes of sidebands $\frac{f_{d2}}{2} \pm n(f_{d1} - f_{d2}); n = 1,2$ do not increase with $S(f_{d2})$. This means that the increase in drive level of $S(f_{d2})$ is only utilized in boosting the excitation level of $\frac{f_{d2}}{2}$ instead of FWM interaction involved in sideband generation. Hence, for the frequency comb about $\frac{f_{d2}}{2}$, the amplitudes of sub-harmonic tone $\frac{f_{d2}}{2}$ and side-bands $\frac{f_{d2}}{2} \pm n(f_{d1} - f_{d2}); n = 1,2$ are influenced by $S(f_{d2})$ and $S(f_{d1})$ respectively. Along the similar lines, for the frequency comb about $\frac{f_{d1}}{2}$, the amplitudes of sub-harmonic tone $\frac{f_{d1}}{2}$ and side-bands $\frac{f_{d1}}{2} \pm n(f_{d1} - f_{d2}); n = 1,2$ are influenced by $S(f_{d1})$ and $S(f_{d2})$ respectively.

Figures 2B-2D presents the qualitative picture of frequency spectra associated with each of the regimes. While the R1 and R3 are represented by a well-defined frequency comb, the frequency spectrum corresponding to R2 is ill-structured. Now, for variable drive levels $S(f_{d1}) = -5, 0, 5\ \&\ 15\ dBm$, we recorded the drive levels $S(f_{d2}) = T_{R1 \rightarrow R2}, T_{R2 \rightarrow R3}$ corresponding to the transitions from R1 to R2 and R2 to R3. As it can be seen from the figure 2E, both $T_{R1 \rightarrow R2}\ \&\ T_{R2 \rightarrow R3}$

monotonously reduces with the increase in $S(f_{d1})$. This means, for greater $S(f_{d1})$, it is easy for the transition from $\frac{f_{d1}}{2} \pm n(f_{d1} - f_{d2})$ to $\frac{f_{d2}}{2} \pm n(f_{d1} - f_{d2})$ to occur. However, if the higher signal levels of $f_{d1}$ and $f_{d2}$ solely favour frequency combs corresponding to the respective sub-harmonic frequencies i.e. $\frac{f_{d1}}{2}$ and $\frac{f_{d2}}{2}$, the transition from $\frac{f_{d1}}{2} \pm n(f_{d1} - f_{d2})$ to $\frac{f_{d2}}{2} \pm n(f_{d1} - f_{d2})$ should be more difficult for greater $S(f_{d1})$. Despite this specific nature, we have know that there is a positive dependence of $S(f_{d1})$ on the strength of sidebands $\frac{f_{d2}}{2} \pm n(f_{d1} - f_{d2})$. This means that greater $S(f_{d1})$ can also favour frequency comb $\frac{f_{d2}}{2} \pm n(f_{d1} - f_{d2})$. Hence, from a general perspective, the thresholds $T_{R1 \to R2}$ & $T_{R2 \to R3}$ are thus controlled by the interplay between both these individual processes. However, in our specific experiments, greater $S(f_{d1})$ has been observed to collectively favour $\frac{f_{d2}}{2} \pm n(f_{d1} - f_{d2})$.

In summary, this paper reports the observation of frequency transitions in phononic FWM process mediated by parametric resonance. While these present an initial indication of complexity concomitant with two tone excitation specific to FWM, the emergent responses of FWM for multiple tone excitations are to be understood through follow-up theoretical and experimental studies. Similar to the observed transitions in FWM, a transition phase between TWM and FWM may also be equally existent. These frequency transitions further present a framework that enables amplitude dependent excitations of multiple frequency states which can be potentially useful for frequency-based computing applications [10].

**Acknowledgements**

Funding from the Cambridge Trusts is gratefully acknowledged.

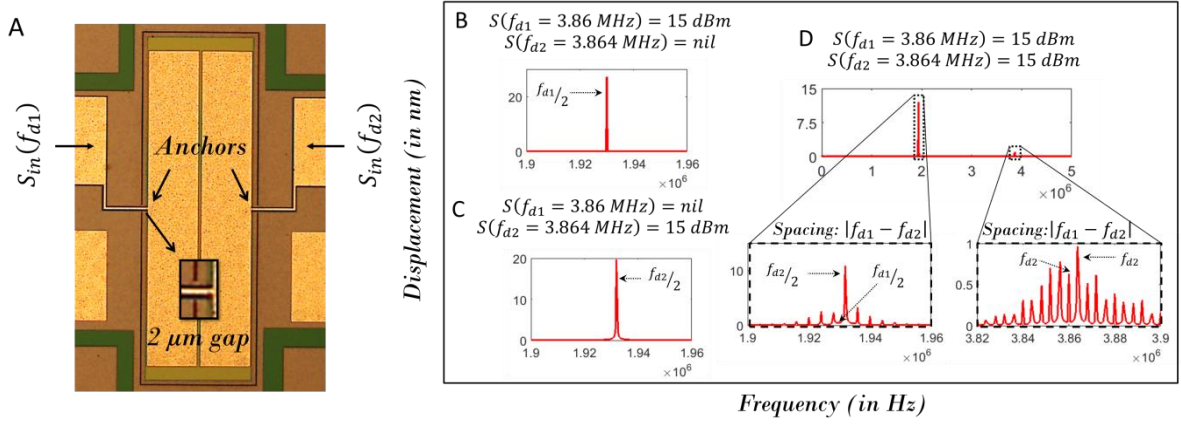

Figure 1: **Absence of simultaneous phononic four-wave mixing (FWM) processes.** A: Signals $S(f_{d1})$ and $S(f_{d2})$ are applied on a free-free beam microstructure; B and C: Parametric excitation of tones at $\frac{f_{d1}}{2}$ and $\frac{f_{d2}}{2}$ for the drive conditions $S(f_{d1} = 3.86\ MHz) = 15\ dBm; S(f_{d2} = 3.864\ MHz) = nil$ and $S(f_{d1} = 3.86\ MHz) = nil; S(f_{d2} = 3.864\ MHz) = 15\ dBm$ respectively; D: The excitation of frequency combs is only about $f_{d2}$ & $\frac{f_{d2}}{2}$.

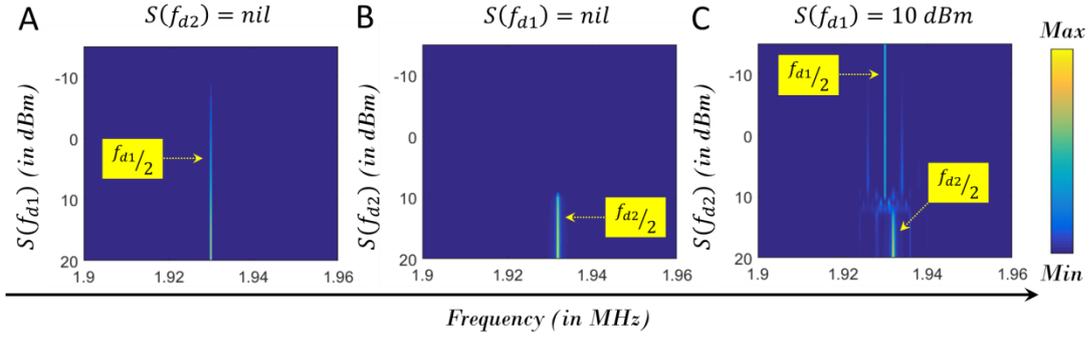

Figure 2: **Frequency transitions in phononic FWM.** A-C: The displacement maps for the drive conditions $(f_{d1} = 3.86\ MHz) = -15 - 20\ dBm$; $S(f_{d2} = 3.864\ MHz) = nil$, $S(f_{d1} = 3.86\ MHz) = nil$; $S(f_{d2} = 3.864\ MHz) = -15 - 20\ dBm$ and $S(f_{d1} = 3.86\ MHz) = 10\ dBm$; $S(f_{d2} = 3.864\ MHz) = -15 - 20\ dBm$ respectively.

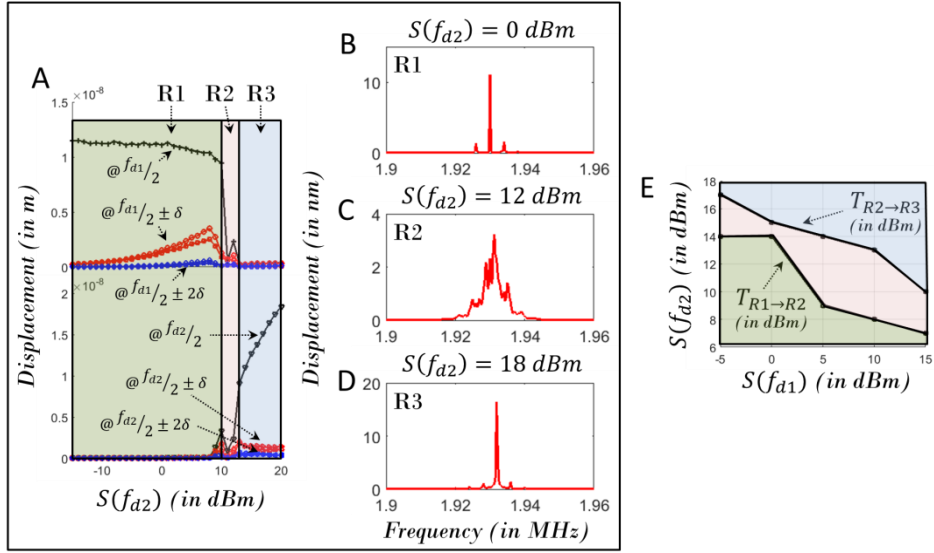

Figure 3: **Drive level dependence of frequency transition.** A: The displacement amplitudes at $\frac{f_{d1}}{2} = 1.93\ MHz$; $\frac{f_{d1}}{2} + (f_{d1} - f_{d2}) = 1.934\ MHz$; $\frac{f_{d1}}{2} - (f_{d1} - f_{d2}) = 1.926\ MHz$; $\frac{f_{d1}}{2} + 2(f_{d1} - f_{d2}) = 1.938\ MHz$; $\frac{f_{d1}}{2} - 2(f_{d1} - f_{d2}) = 1.922\ MHz$; $\frac{f_{d2}}{2} = 1.932\ MHz$; $\frac{f_{d2}}{2} + (f_{d1} - f_{d2}) = 1.936\ MHz$; $\frac{f_{d2}}{2} - (f_{d1} - f_{d2}) = 1.928\ MHz$; $\frac{f_{d2}}{2} + 2(f_{d1} - f_{d2}) = 1.940\ MHz$ and $\frac{f_{d2}}{2} - 2(f_{d1} - f_{d2}) = 1.924\ MHz$. This indicates the three different regimes R1, R2 and R3; B-C: The representative frequency spectra corresponding to R1, R2 and R3 for the drive conditions $(f_{d1} = 3.86\ MHz) = 10\ dBm$; $S(f_{d2} = 3.864\ MHz) = 0\ dBm$, $S(f_{d1} = 3.86\ MHz) = 10\ dBm$; $S(f_{d2} = 3.864\ MHz) = 12\ dBm$ and $S(f_{d1} = 3.86\ MHz) = 10\ dBm$; $S(f_{d2} = 3.864\ MHz) = 18\ dBm$ respectively; E: The drive levels of $S(f_{d2} = 3.864\ MHz)$ at which transition from R1 to R2 and R2 to R3 occurs for different drive levels of $S(f_{d1} = 3.86\ MHz)$.